\documentclass[12pt,aps,preprint,nofootinbib,superscriptaddress,nobalancelastpage]{revtex4}

\usepackage{mathrsfs}
\usepackage{amsmath, amsthm, amssymb}
\usepackage{color}
\usepackage{graphicx}
\usepackage{verbatim}

\usepackage{hyperref}	
\hypersetup{colorlinks,bookmarksopen,bookmarksnumbered,citecolor=blue,
linkcolor=blue,pdfstartview=FitH,urlcolor=blue}

\newcommand{\be}{\begin{equation}}
\newcommand{\ee}{\end{equation}}
\newcommand{\bea}{\begin{eqnarray}}
\newcommand{\eea}{\end{eqnarray}}

\newcommand{\fig}{Fig.}

\newcommand{\Ref}{Ref.}
\newcommand{\Refs}{Refs.}

\newcommand{\ie}{\emph{i.e.}}

\newcommand{\bld}[1]{{\boldsymbol{#1}}}

\newcommand{\ESS}{ESS$\nu$SB}
\newcommand{\ESSt}{ESS$\bld{\nu}$SB}

\newcommand{\Cerenkov}{Cherenkov}

\newcommand{\kt}{kt}

\begin{document}

\preprint{FTUAM-14-24, IFT-UAM/CSIC-14-059, NORDITA-2014-81}

\title{Searching for sterile neutrinos at the \ESSt}

\author{Mattias Blennow}
\email{emb@kth.se}
\affiliation{Department of Theoretical Physics, School of Engineering Sciences, KTH Royal Institute of Technology, AlbaNova University Center, 106 91 Stockholm, Sweden}
\author{Pilar Coloma}
\email{pcoloma@vt.edu}
\affiliation{Center for Neutrino Physics, Virginia Tech, Blacksburg, VA 24061, USA}
\author{Enrique Fernandez-Martinez}
\email{enrique.fernandez-martinez@uam.es}
\affiliation{Departamento de F\'isica Te\'orica, Universidad Aut\'onoma de Madrid, Cantoblanco E-28049 Madrid, Spain}
\affiliation{Instituto de F\'isica Te\'orica UAM/CSIC,
 Calle Nicol\'as Cabrera 13-15, Cantoblanco E-28049 Madrid, Spain}

\begin{abstract}
The \ESS\ project is a proposed neutrino oscillation experiment based on the European Spallation Source with the search for leptonic CP as its main aim. In this letter we show that a near detector at around 1~km distance from the beamline is not only very desirable for keeping the systematic errors affecting the CP search under control, but would also 
provide a significant sensitivity probe for sterile neutrino oscillations in the region of the parameter space favored by the long-standing LSND anomaly. We find that the effective mixing angle $\theta_{\mu e}$ can be probed down to $\sin^2(2\theta_{\mu e}) \simeq 2(8)\cdot 10^{-3}$ at $5\sigma$ assuming 15~\% bin-to-bin (un)correlated systematics.
\end{abstract}

\maketitle

\section{Introduction}
\label{sec:intro}

Since the first evidence for neutrino oscillations in 1998, the experimental progress in the field has been remarkable. Today the standard three-flavor paradigm of neutrino oscillations is well tested and most of the parameters it contains are known to high precision~\cite{GonzalezGarcia:2012sz}. This framework includes six parameters out of which three are mixing angles, two are independent mass squared differences and one is a, yet unknown, CP violating phase. However, there are a number of experimental anomalies that do not fit into this picture. The observation of an excess of $\bar{\nu}_e$ events at $3.8\sigma$ at the LSND experiment~\cite{Aguilar:2001ty} could be explained by introducing one or more additional \emph{sterile} neutrino states, with an additional mass squared difference of $\Delta m^2_{41} \sim \mathcal{O}(\textrm{eV}^2)$ in order to allow for $\nu_\mu \to \nu_e$ oscillations in the observed range of $L/E$. However, KARMEN did not observe any signal for similar values of $L/E$~\cite{Armbruster:2002mp}.  The dedicated MiniBooNE experiment~\cite{AguilarArevalo:2007it,AguilarArevalo:2010wv,Aguilar-Arevalo:2013pmq} has to this day not confirmed or disproved the LSND anomaly. On the other hand, $\nu_e$~\cite{Bahcall:1994bq,Giunti:2006bj,Giunti:2010wz,Giunti:2010zu} and $\bar\nu_e$~\cite{Mueller:2011nm,Mention:2011rk,Huber:2011wv} disappearance experiments seem to observe a deficit which would be compatible with sterile neutrino oscillations in the eV$^2$ range, while no deficit at all has been observed in any of the $\nu_\mu$ disappearance experiments~\cite{Dydak:1983zq,Maltoni:2007zf}. Overall, global analyses show strong tension between different data sets~\cite{Kopp:2013vaa,Conrad:2013mka,Giunti:2013aea} and further experimental input will be needed to clarify the situation. 

The European Spallation Source $\nu$-Beam~\cite{Baussan:2012cw,Baussan:2013zcy} (\ESS) is a proposal for a neutrino oscillation experiment based upon the accelerator facilities of the future European Spallation Source~(ESS) and optimized for the goal of a high significance search for CP violation. The relatively large value of $\theta_{13}$ recently discovered~\cite{Adamson:2011qu,Abe:2011sj,An:2012eh,Ahn:2012nd,Abe:2012tg} guarantees a comparatively high signal, which implies that the bottleneck of CP violation searches for the next generation of neutrino oscillation facilities will typically be systematics errors rather than statistics or backgrounds. The effect of such systematic uncertainties can be alleviated in two ways. If the statistics is high enough, placing the detector close to the second oscillation maximum enhances the relative importance of the CP-violating component of the oscillation probability, increasing the sensitivity of the facility and reducing the impact of systematic errors~\cite{Coloma:2011pg}. Indeed, the optimization of the \ESS\ showed a clear preference for this location which was adopted as its baseline design. The second way of controlling the negative impact of systematic uncertainties on the CP violation search is the inclusion of a near detector at $\sim 1$~km distance from the source in order to calibrate the neutrino flux and cross sections. However, a near detector could also play a role in testing new physics in the neutrino sector. Indeed, since the neutrino flux at the \ESS\ would be peaked in the $\mathcal O(0.2$~GeV) range, a near detector placed at a baseline of 1~km would be ideal for testing oscillations of sterile neutrinos in the parameter range indicated by the LSND anomaly. Furthermore, the high power of the beam (5~MW) would enable the possibility to test very small values of the oscillation probability with a very high significance.

In the following we will study the sensitivity of the \ESS\ setup, using only a near detector placed at 1~km from the source, to oscillations in the $\nu_\mu \to \nu_e$ appearance channel. In order to do so, we will adopt a phenomenological approach where the oscillation probability reads
\begin{equation}
 P(\nu_\mu\to\nu_e) = \sin^2(2\theta_{\mu e}) \sin^2\left(\frac{\Delta m^2_{41} L}{4E}\right).
\label{eq:prob}
\end{equation}
Here, $\theta_{\mu e}$ is an effective mixing angle, $\Delta m^2_{41}\equiv m_4^2 - m_1^2$ is the active-sterile squared mass difference, $L$ is the distance from the source, and $E$ is the neutrino energy. This approach is a reasonable approximation as long as we are dealing with small enough values of $L/E$ such that the active-active neutrino oscillations have not developed significantly yet.

In the following we will investigate the sensitivity of the \ESS\ in two separate scenarios,
\begin{eqnarray*}
 {\rm Case~I:} && {\rm no\ active-sterile\ mixing} \quad {\rm and} \\
 {\rm Case~II:}&& \sin^2(2\theta_{\mu e}) = 1.3\cdot 10^{-2}, \quad \Delta m^2_{41} = 0.42~{\rm eV}^2.
\end{eqnarray*}
Case I corresponds to a scenario without steriles (as suggested by $\nu_\mu$ disappearance experiments) while Case II corresponds to a situation in which the active-sterile oscillation parameters take the best-fit values of \fig~7 of \Ref~\cite{Kopp:2013vaa}, where a global analysis of several $\nu_\mu \rightarrow \nu_e$ and $\bar\nu_\mu \rightarrow\bar\nu_e$ experiments was performed.

\section{Setup and simulation details}

The GLoBES software~\cite{Huber:2004ka,Huber:2007ji} has been used in order to simulate the \ESS\ experiment. Since the flux peak would be located around 250~MeV, a distance to the detector of 1~km would be needed in order to match the maximum of the oscillation, assuming a squared mass splitting $\Delta m^2_{41} \sim \mathcal{O}(1~\textrm{eV})$ (as LSND~\cite{Aguilar:2001ty} and MiniBooNE~\cite{AguilarArevalo:2007it,AguilarArevalo:2010wv,Aguilar-Arevalo:2013pmq} results seem to indicate). Based on the planned far detector solution, we take a 1~\kt\ water \Cerenkov\ detector as the near detector and we assume it to be identical to the far detector in terms of efficiencies and background rejection capabilities. The response of the detector has been simulated using migration matrices, signal and background rejection efficiencies from \Ref~\cite{Agostino:2012fd}.

Neutrino fluxes have been explicitly simulated for a near detector placed at 1~km from the source~\cite{nikos}. A beam power of 5~MW and $ 1.7\times 10^7$ operating seconds per year with 2.5~GeV protons, as for the long baseline experiment~\cite{Baussan:2013zcy}, is assumed. In order to respect the configuration of the long baseline experiment (which is optimized for CP violating searches) we keep the same ratio of neutrino (2 years) and antineutrino (8 years) running times\footnote{We note, however, that in order to optimize the experiment to search for a sterile neutrino, a different ratio may be desirable, in particular taking into account the different size of neutrino and antineutrino cross sections. See also \Ref~\cite{Agarwalla:2014tpa} for optimizations to standard oscillation physics.}. In absence of oscillations, this would yield a total of 4.00$\times 10^6$ (2.29$\times 10^6$) unoscillated $\nu_\mu$ ($\bar\nu_\mu$) events at the detector during a 10~year life-time of the experiment, where efficiencies have already been accounted for. On the other hand, we find that the expected total number of background events in the $\nu_e$ and $\bar\nu_e$ channels would be $\sim 34400$ and $\sim 23100$, respectively, assuming identical background rejection capabilities as in \Refs~\cite{Baussan:2013zcy,Agostino:2012fd}. The largest contributions to the background event rates would come from $\nu_\mu$ mis-identified as $\nu_e$, and from neutral current~(NC) events mis-identified as charged-current (CC) events.

The neutrino fluxes at the \ESS\ will be peaked within a relatively narrow energy range, see \Ref~\cite{Baussan:2013zcy}. The energy resolution for a water \Cerenkov\ detector would limit the size of the energy bins to $\sim 100$~MeV, implying that only a few energy bins would contain a sizable number of events. However, we have found that the sensitivity of the setup would be compromised if the analysis was done as a counting experiment only (without energy bin separation). The main reason for this is that the signal and background spectra show a different dependence with neutrino energy, which implies that different energy bins will have a different signal-to-background ratio. Therefore, we compute a binned $\chi^2$ in neutrino energy, with nine $100$~MeV bins between 0.1 and 1~GeV. 

Given the high statistics at the near detector, the performance will be limited by systematics. Thus, we will show our results in two scenarios with a different implementation of systematic errors. For the more optimistic option an overall 15~\% uncertainty, uncorrelated between the different channels and between the signal and background components but fully correlated among the bins of each channel, is considered. For the more conservative option, on the other hand, we allow uncorrelated systematics among the different energy bins in each channel, so as to account for uncertainties in the shape of the fluxes and cross sections. However, in order to accommodate an overall normalization shift for the signal in this case, we would be effectively multiplying the penalty in terms of $\chi^2$ by the number of bins with respect to the more optimistic case. Therefore, in this conservative scenario we also add an additional 15~\% normalization uncertainty which is correlated among all bins in order to avoid this behaviour. A modified version of the GLoBES software as in \Ref~\cite{Coloma:2012ji} was used for this implementation of the systematics. A pull-term corresponding to each uncertainty is added to the $\chi^2$, and the result is then marginalized over the nuisance parameters to search for the minimum. The procedure is then repeated for all points in the $(\sin^2(2\theta_{\mu e}), \Delta m^2_{41})$ parameter space, and contours for equal values of the $\chi^2$ are drawn.

Finally, since in this work we are mainly interested in the performance of the \ESS\ with regard to sterile neutrino oscillations, the far detector is not considered in our simulations.

\section{Results}

\begin{figure}
  \includegraphics[width=0.45\textwidth]{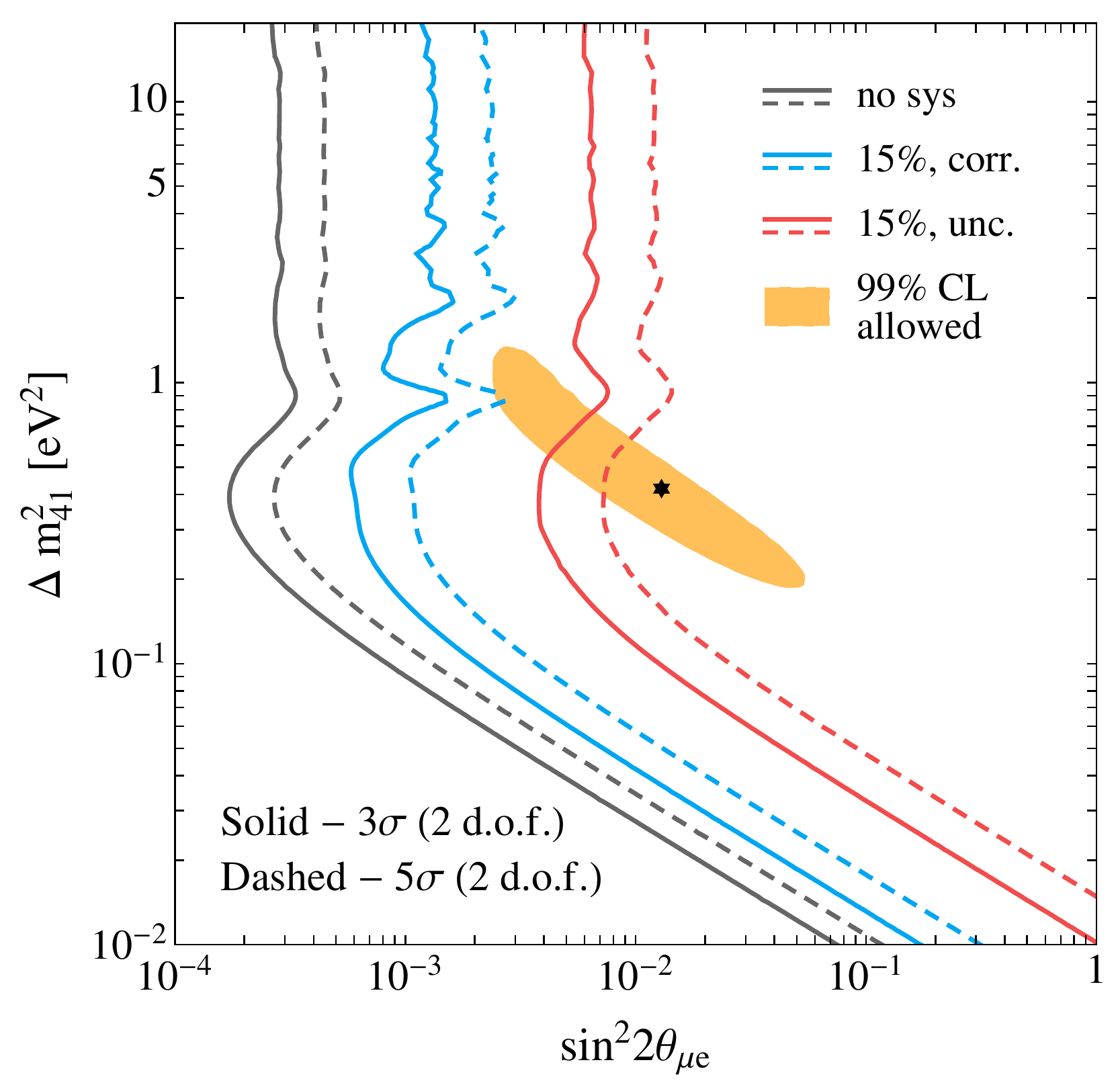}\includegraphics[width=0.45\textwidth]{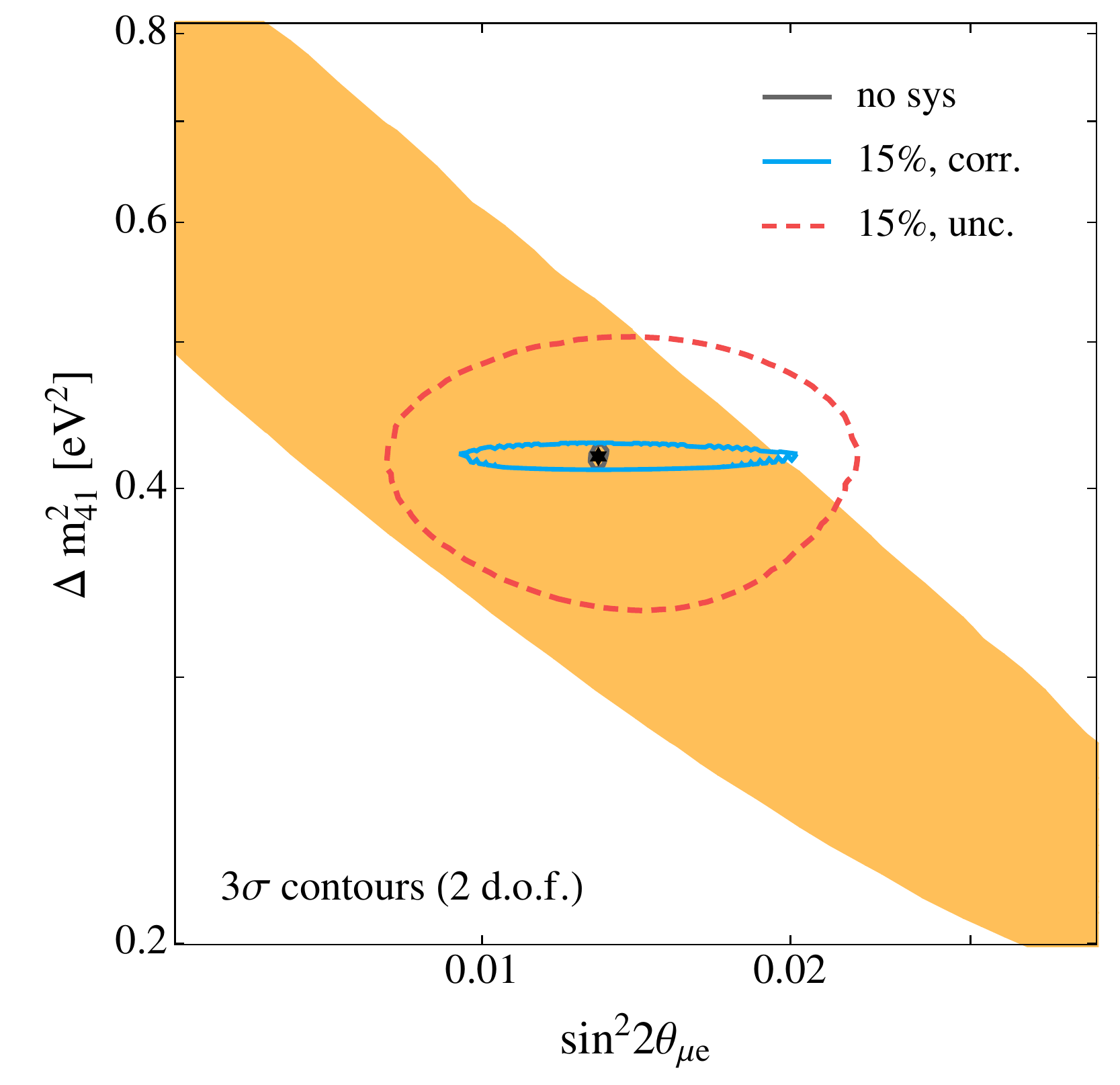}
  \caption{\label{fig:nosignal} Left: The 3 and 5$\sigma$ exclusion contours for totally correlated and totally uncorrelated bin-to-bin systematics of $15 \%$. The no-systematics limit is also shown. The shaded region corresponds to the allowed region at 99\% CL obtained from a global fit to $\nu_e$ disappearance and $\nu_\mu \to \nu_e $ appearance experiments, taken from \fig~7 of \Ref~\cite{Kopp:2013vaa}. Right: The expected confidence regions for Case~II at 3~$\sigma$ for totally correlated and totally uncorrelated bin-to-bin systematics of $15 \%$, as indicated in the legend.}
\end{figure}

The left panel of Fig.~\ref{fig:nosignal} shows the expected 3 and $5\sigma$ exclusion limits of the \ESS\ under the assumption of no active-sterile neutrino mixing (Case~I).
As expected from the baseline length of 1~km and typical neutrino energy of $\sim 0.3$~GeV, the best sensitivity is achieved around $\Delta m^2_{41} \simeq 0.3$~eV$^2$. Below this value, the sensitivity in $\sin^2(2\theta_{\mu e})$ quickly deteriorates as the oscillations do not have sufficient time to develop. On the other hand, for larger values of the mass splitting the oscillations enter the averaged regime. For comparison with the current experimental situation, the allowed region at 99~\%~CL from a global fit to $\nu_e$ disappearance and $\nu_\mu \to \nu_e$ appearance experiments is also shown, see \Ref~\cite{Kopp:2013vaa}.
As seen in the figure, the final results will be dominated by systematic errors. Assuming bin-to-bin correlated 15~\% systematic errors the currently allowed region could be completely excluded at 5$\sigma$ from this measurement alone, while the sensitivity in the more conservative scenario of completely uncorrelated errors is somewhat more limited. However, we find that also in this scenario most of the preferred region, including the best fit, is covered with 5$\sigma$ significance. Finally, we also show in this panel that in the limit of no systematic uncertainties values of $\sin^2(2\theta_{\mu e})$ down to $\sim 10^{-4}$ could be probed.   

The right panel of Fig.~\ref{fig:nosignal}, on the other hand, shows the expected $3\sigma$ confidence regions under the assumption of active-sterile neutrino mixing with oscillation parameters corresponding to the best-fit values from \Ref~\cite{Kopp:2013vaa} (Case~II). From this figure it can be seen that both the mass squared difference and the mixing angle would be very well determined by the \ESS\ experiment in case of a positive signal. When assuming correlated systematics, the error on the mass squared difference improves noticeably. The main reason for this is the large number of events in each bin, combined with the fact that the value of $L/E$ for the setup considered in this work would match very well the first oscillation maximum for $\Delta m_{41}^2\sim 0.4$~eV. Thus, small changes in $\Delta m_{41}^2$ result in large corrections to  the expected number of events in each bin, providing a good sensitivity to $\Delta m_{41}^2$. On the other hand, the systematic on the overall strength of the signal is directly correlated to the value of $\sin^2(2\theta_{\mu e})$ and therefore is more difficult to constrain under the assumption of fully correlated systematics. 

\section{Summary and conclusions}

In this letter we have discussed the possibility of using a near detector at the recently proposed \ESS\ neutrino oscillation experiment in order to look for active-sterile neutrino oscillations in the range indicated by the LSND anomaly, $\Delta m^2_{41} \sim 1$~eV$^2$ and $\sin^22\theta_{\mu e}\sim 10^{-2}$. Our study is based on the performance of a 1~\kt\ near water \Cerenkov\ detector at a distance of 1~km from the source. Under the assumption of no active-sterile neutrino mixing, we find that the \ESS\ setup would be able to completely exclude the currently allowed region from a global fit to $\nu_e$ disappearance and $\nu_\mu \to \nu_e$ appearance data with a confidence of 5$\sigma$, assuming bin-to-bin correlated systematical errors at the 15~\% level. Even in the more conservative scenario of fully uncorrelated bin-to-bin systematics, most of the preferred region would also be covered with 5$\sigma$ significance. On the other hand, if active-sterile mixing takes place with oscillation parameters in the range currently favored by global analyses, the \ESS\ experiment would be able to pinpoint their values with extremely good accuracy. 

Finally it should also be noted that, in order not to interfere with the main goal of the \ESS\ (\ie, the discovery of leptonic CP violation), no optimization for sterile neutrino searches has been performed in this work. Therefore, if a signal of the existence of sterile neutrinos was to be found, some room for improvement over the results obtained here may be possible.

\begin{acknowledgments}
The authors would like to thank Nikolaos Vassilopoulos for providing the \ESS\ fluxes at a distance of 1~km, and the authors of \Ref~\cite{Kopp:2013vaa} for providing the curves for their global fit allowed regions for comparison.

This work has been supported by the G\"oran Gustafsson Foundation~[MB], and by the U.S. Department of Energy under award number \protect{DE-SC0003915} [PC]. EFM acknowledges financial support by the European Union through the FP7 Marie Curie Actions CIG NeuProbes (PCIG11-GA-2012-321582) and the ITN INVISIBLES (PITN-GA-2011-289442), and the Spanish MINECO through the ``Ram\'on y Cajal" programme (RYC2011-07710) and through the project FPA2009-09017. We also thank the Spanish MINECO (Centro de excelencia Severo Ochoa Program) under grant \protect{SEV-2012-0249} as well as the Nordita Scientific program ``News in Neutrino Physics'', where this work was initiated.
\end{acknowledgments}


\end{document}